\documentstyle[aps,multicol,psfig,amsfonts]{revtex}
\newcommand{\ds}{\rule[-1.5mm]{0mm}{4.5mm}\displaystyle}
\renewcommand{\bar}[1]{\overline{#1}}
\begin{document}   
\newcommand{\bra}{\langle}
\newcommand{\ket}{\rangle}
\bibliographystyle{prsty}
\widetext  
\title{  
Eigenvector statistics
in non-Hermitian random matrix ensembles}
\author{J. T. Chalker and B. Mehlig}
\address{Theoretical Physics, University of Oxford,
         1 Keble Road, OX1 3NP, United Kingdom}  
\date{\today}
\maketitle{ } 
\begin{abstract}  
We study statistical properties of the eigenvectors of 
non-Hermitian random matrices, concentrating on 
Ginibre's complex Gaussian ensemble, in which
the real and imaginary parts of each element of 
an $N \times N$ matrix, $J$,
are independent random variables.
Calculating ensemble averages based on the quantity
$\langle L_\alpha | L_\beta \rangle\,
                  \langle R_\beta | R_\alpha\rangle\,$,
where $\langle L_\alpha |$ and $| R_\beta \rangle\,$
are left and right eigenvectors of $J$,
we show for large $N$ that eigenvectors
associated with a pair of eigenvalues
are highly correlated if the two 
eigenvalues lie close in the complex plane.
We examine consequences of these correlations 
that are likely to be important in physical applications.
\end{abstract}   
\begin{multicols}{2}
An understanding of statistical properties of ensembles 
of random matrices has proved useful in many different 
areas of physics \cite{meh67}.
Because the first application
of random matrix theory, and one
still of great importance, was to 
represent the Hamiltonian of a non-integrable quantum system,
the early work of Wigner, Dyson and others focussed on
ensembles of real symmetric or complex Hermitian
matrices. Eigenvector distributions in these ensembles
are of limited interest, being determined by the Haar 
measure on the group that leaves the ensemble invariant. 
Instead, the concern is mainly 
with eigenvalue correlations, about which a great deal is now known \cite{meh67}.

More recently, the spectral properties of random non-Hermitian 
operators have
attracted attention in a variety of contexts,
including: neural network dynamics \cite{som88};
the quantum mechanics of open systems \cite{haa92};
the statistical mechanics of flux lines in superconductors
with columnar disorder \cite{hat96,efe97a,gol97,hat98b,mud98}; 
classical diffusion in
random media \cite{cha97}; and biological growth problems \cite{nel98}.
The corresponding ensembles of real asymmetric and 
general complex matrices were first studied by
Ginibre \cite{gin65}, Girko \cite{gir85},
and Sommers and co-workers \cite{som88,leh91}.
The eigenvalues in these ensembles are, of course, not restricted to
the real axis, but rather distributed over an area in the complex plane.
Their density and correlations have been
investigated in considerable detail 
\cite{som88,gin65,gir85,leh91,fyo97}.

By contrast, {\it eigenvector} statistics in non-Hermitian random matrix ensembles have not, so far as we know, previously been examined. The existence of distinct sets of left and right eigenvectors means that invariance of the ensemble, under $O(N)$ or $U(N)$ transformations as appropriate, is a rather weak constraint on the joint eigenvector distribution: it generates no information on the relative orientations of the two sets of vectors. We 
show in this paper that there are, in fact, remarkable correlations between left and right eigenvectors. These correlations are likely to be important in physical applications of non-Hermitian random matrix ensembles. 
We illustrate their significance by discussing two consequences,
involving extreme sensitivity of spectra to perturbations, and transients in 
time-evolution, which are 
recognised in other contexts as typical of non-normal operators \cite{tref93,far96,tref97}.

We consider, following Ginibre \cite{gin65}, the
Gaussian ensemble of general
complex $N\times N$ matrices, $J$, having independent
matrix elements, $J_{kl}$, distributed with
probability
\begin{eqnarray}
\label{eq:ensemble}
P(J)\, dJ \propto \,\exp(-N\mbox{\small Tr}\left[J J^{\dagger}\right]) \, 
\prod_{k,l=1}^N \! dJ_{kl}^\prime \, d J_{kl}^{\prime\prime}
\end{eqnarray}
where $J_{kl} = J_{kl}^\prime + i J_{kl}^{\prime\prime}$,
with $J_{kl}^\prime$ and $J_{kl}^{\prime\prime}$
real.
Denoting ensemble averages by  $\langle\cdots\rangle$ and complex conjugation 
with an overbar,
the only non-zero cumulant  of $J$ is
$\langle J_{kl}\,\bar{J}_{kl}\rangle = 1/N$.

The eigenvalues, $\lambda_{\alpha}$, of $J$ are distributed in 
the complex plane with, in the limit $N \to \infty$, 
constant density inside a disc
of unit radius, centered on the origin.
They are non-degenerate with probability one, 
and in this case the left and right eigenvectors, $\langle L_\alpha|$
and $|R_\alpha\rangle$,
which satisfy
\begin{eqnarray}
\label{eq:eveq}
J\,|R_\alpha\rangle   &=& \lambda_\alpha\,|R_\alpha\rangle\,,\\
\langle L_\alpha|\,J &=& \langle L_\alpha|\,\lambda_\alpha\,.
\nonumber
\end{eqnarray}
form two complete, biorthogonal sets, and can be normalised so that
\begin{equation}
\label{eq:biorth}
\langle L_\alpha|R_\beta\rangle = \delta_{\alpha\beta}\,.
\end{equation}
We indicate Hermitian conjugates of vectors in the usual way, so that, for example, $|L_\alpha\rangle$ satisfies $J^{\dagger}\,|L_\alpha\rangle= \bar{\lambda}_\alpha\,|L_\alpha\rangle$.

We investigate eigenvector correlations mainly by calculating ensemble averages of combinations of scalar products. Noting that 
Eqs.\,(\ref{eq:eveq}) and (\ref{eq:biorth})
are invariant under a scale transformation
$|R_\alpha\rangle \rightarrow \zeta_\alpha\,|R_\alpha\rangle$
and $\langle L_\beta| \rightarrow \langle L_\beta |\, {\zeta}_\beta^{-1}$,
one recognises that only those combinations invariant under this transformation should be considered.
The simplest such combination, involving
two eigenvectors, is fixed by Eq. (\ref{eq:biorth});
the simplest non-trivial quantity
is thus the matrix of overlaps
\begin{equation}
\label{eq:defO}
O_{\alpha\beta} = \langle L_\alpha | L_\beta \rangle\,
                  \langle R_\beta  | R_\alpha\rangle\,.
\end{equation}
and we shall focus on this throughout the paper.
It is convenient to define 
local averages of diagonal and off-diagonal 
elements of the overlap matrix,
\begin{eqnarray}
\label{eq:Odiag}
O(z) &=&  \Big\langle\frac{1}{N}\sum_\alpha O_{\alpha\alpha}
\,\delta(z-\lambda_\alpha)\Big\rangle\,,\\
\label{eq:Ooff}
O(z_1,z_2) &=&\Big\langle \frac{1}{N}\sum_{\alpha\neq\beta} O_{\alpha\beta}
\,\delta(z_1-\lambda_\alpha) \,\delta(z_2-\lambda_\beta)
\Big\rangle\,.
\end{eqnarray}
Correspondingly,
the density of states is defined as
$d(z) = \big\langle N^{-1}\sum_\alpha \delta(z-\lambda_\alpha)\big\rangle$.
In the limit $N\rightarrow\infty$, $d(z) = \pi^{-1}$
for $|z|<1$ and $d(z) = 0$ otherwise \cite{gin65}.

We have been able to obtain exact expressions
for $O(z_1)$ and $O(z_1,z_2)$. 
For $N \gg 1$, $|z_1-z_2| \neq 0$
and $|z_1|,|z_2| <1 $ these simplify to
\begin{eqnarray}
\label{eq:largeNdiag}
O(z_1)    &=& \frac{N}{\pi} \left(1-|z_1|^2\right)\,,\\
\label{eq:largeNoff}
O(z_1,z_2)&=&-\frac{1}{\pi^2}
\frac{\ds 1-z_1\bar{z}_2}{\ds |z_1-z_2|^4}\,.
\end{eqnarray}
For $|z_1|,|z_2| \geq 1$, 
both densities vanish as $N\rightarrow\infty$.
To display the form of $O(z_1,z_2)$ 
as $|z_1 - z_2| \to 0$, it is necessary to express  $|z_1 - z_2|$
in units of the separation between adjacent eigenvalues,
introducing
$z_+ = (z_1+z_2)/2$ and
$\omega = \sqrt{N}\,(z_1-z_2)$. We obtain,
for $|z_+ | < 1$, $\omega \ll\sqrt{N}$
and $N\gg 1$
\begin{eqnarray}
\label{eq:np}
O(z_1,z_2)&=&
-N^2\frac{1-|z_+|^2}{\pi^2|\omega|^4}
\left(1-\big(1+|\omega|^2)\,{\rm e}^{-|\omega|^2}
\right)\,.
\end{eqnarray}
Eqs.\,(\ref{eq:largeNdiag}) -- (\ref{eq:np}) constitute our main results.
Before outlining our derivation, we discuss their significance.

First, we stress the dramatic difference between the behaviour of $O_{\alpha\beta}$ in this general complex ensemble and its behaviour in the case of Hermitian matrices, for which $O_{\alpha\beta}=\delta_{\alpha\beta}$.
The fact that, by contrast, $O_{\alpha\alpha} \sim N$ in the non-Hermitian ensemble
can be understood as the behaviour 
which results if $\langle L_\alpha |$
and $| R_\alpha \rangle$ are independent random vectors, subject to the normalisation of Eq.\,(\ref{eq:biorth}). Moreover, large values for the diagonal elements of the matrix $O_{\alpha\beta}$ must be accompanied by some large (or many small)
off-diagonal elements, since the two are linked by a sum rule that follows from completeness,
\begin{equation}
\label{eq:sumrule}
\sum_{\alpha} O_{\alpha\beta} = 1\,.
\end{equation}
Indeed, Eq. (\ref{eq:Ooff}) implies
\begin{equation}
\label{eq:Oab}
O_{\alpha\beta} \sim O(z_1,z_2)\big/\Big\langle \frac{1}{N}\sum_{\mu\neq\nu}
\delta(z_1-\lambda_\mu)\,\delta(z_2-\lambda_\nu)\Big\rangle
\end{equation}
and hence, from Eq. (\ref{eq:np}),
$O_{\alpha\beta} \sim - N$
 if $\lambda_{\alpha}$ and $\lambda_{\beta}$ are neighbouring 
 eigenvalues in the complex plane, so that (typically) $\omega \sim 1$.

An immediate consequence of large values of $O_{\alpha\alpha}$
is that the spectrum of $J$ for a given realisation has extreme sensitivity to 
perturbations.
Consider
for definiteness 
$J = \cos(\theta) J_1 + \sin(\theta) J_2$, where $\theta$ is real
and $J_1$ and $J_2$ 
are both drawn independently from the ensemble of Eq.\,(\ref{eq:ensemble}),
so that $J$ moves through the ensemble as $\theta$ varies.
Then
\begin{equation}
|\partial \lambda_\alpha/\partial \theta|^2
= | \langle L_\alpha| \partial J/ \partial \theta |R_\alpha\rangle|^2\,.
\end{equation}
Performing the ensemble average, and using Eq. (\ref{eq:largeNdiag})
one obtains for the mean square eigenvalue velocity
\begin{equation}
\label{eq:sens}
\left\langle |\partial \lambda/\partial \theta|^2\right\rangle 
= \frac{1}{\pi}\big(1-|\lambda|^2\big) \,.
\end{equation}
This result should be contrasted in magnitude 
with the analogous one for Hermitian matrices \cite{wilk89}. 
Let $H= \cos(\theta) H_1 + \sin(\theta) H_2$, where $H_1$ and $H_2$ are
complex Hermitian $N \times N$ matrices, drawn independently from the Gaussian unitary ensemble, in which the non-zero cumulants are
$\langle H_{kl}\,H_{lk}\rangle \equiv \langle H_{kl}\,\bar{H}_{kl}\rangle = 1/N$. Let $E$ be an eigenvalue of $H$.
Then for $N \rightarrow \infty$, $-1 \leq E \leq 1$ and
$\langle [\partial E/\partial \theta]^2\rangle 
= 1/N$. 
Thus the eigenvalues of the $N \times N$ random non-Hermitian matrix are ${\cal O}(N)$ times more sensitive to perturbations 
than those of the Hermitian matrix. 
Such sensitivity is known to be a typical property of non-normal operators \cite{tref97}.
Despite the sensitivity of individual eigenvalues to 
perturbations, it is reasonable to expect some stability in 
the structure of the spectrum as a whole, since the perturbations 
considered merely take a random matrix from one realisation to another. 
Such stability arises from the fact that, although the mean square 
velocity of Eq.\,(\ref{eq:sens}) is large for eigenvalues within 
the unit disc, it vanishes as the boundary to the support of 
the spectrum is approached. Conversely, anticipating this stability, 
we have a rationalisation of the fact that, 
from Eqs.\,(\ref{eq:largeNdiag}),(\ref{eq:np}) and (\ref{eq:Oab}),
the amplitudes of the ${\cal O}(N)$ contributions to $O_{\alpha\alpha}$ and 
$O_{\alpha\beta}$ vanish as $|z_1| \to 1$.

The large off-diagonal elements of $O_{\alpha\beta}$ are significant in situations in which $J$ is the generator of evolution in real or imaginary time.
Settings of this type represent one of the main physical applications of non-normal operators \cite{som88,hat96,cha97,nel98}. To be specific, consider a model problem in which
\begin{equation}
\frac{\partial}{\partial t}|u(t)\rangle = (J-1)|u(t)\rangle 
\label{eq:evolution}
\end{equation}
so that
\begin{equation}
\label{eq:ut}
|u(t)\rangle = \sum_{\alpha} |R_\alpha\rangle \,f_t(\lambda_\alpha)\,
  \langle L_\alpha|u(0)\rangle\,,
\end{equation}
with $f_t(\lambda)=\exp([\lambda-1]t)$, where we use $(J-1)$ rather than $J$ in Eq.\,(\ref{eq:evolution}) for convenience, to suppress exponential growth.
Ensemble averaging with $\langle u(0) | u(0) \rangle =1$ leads to
\begin{equation}
\label{eq:ut2}
\big\langle\langle u(t)| u(t)\rangle\big\rangle = 
\Big\langle\frac{1}{N}\sum_{\alpha\beta} O_{\alpha\beta}\,
f_t(\lambda_\alpha)\,
\bar{f}_t(\lambda_\beta)\Big\rangle\,,
\end{equation}
and for $t\gg 1$ and $N\to \infty$ we find \cite{footnote}
\begin{equation}
\big\langle\langle u(t)| u(t)\rangle\big\rangle 
\sim (4\pi t)^{-1/2}\,.
\end{equation}
This behaviour should be compared with the much faster decay that would result from the same spectrum if the eigenvectors were orthogonal. In the same regime, the replacement $O_{\alpha\beta} \to \delta_{\alpha\beta}$ transforms Eq.\,(\ref{eq:ut2}) into
\begin{equation}
\langle \sum_{\alpha} |f_t(\lambda_{\alpha})|^2\rangle  
\sim (4\pi t^3)^{-1/2}\,.
\end{equation}
Thus, eigenvector correlations may be as significant as eigenvalue distributions in determining evolution at intermediate times, a fact of established importance in hydrodynamic stability theory \cite{tref93,far96}.

Finally, it is interesting to ask about, not only the average behaviour of
the overlap matrix, but also its fluctuations. In fact,  $O_{\alpha\beta}$ is typically large if the matrix $J$ has an eigenvalue which is almost degenerate with $\lambda_{\alpha}$ or $\lambda_{\beta}$, and as a result, the probability distribution of  $O_{\alpha\beta}$ has a power-law tail extending to large $|O_{\alpha\beta}|$.
To illustrate this, we consider $N=2$, for which we can calculate 
exactly the probability distribution, $P(O_{\alpha\alpha})$, of a diagonal element of the overlap matrix.
We find
\begin{equation}
\label{eq:dist}
P(O_{\alpha\alpha}) = 4\,\frac{\Theta(O_{\alpha\alpha}-1)}
                           {\,\,(\,2\,O_{\alpha\alpha}-1)^3}\,,
\end{equation}
where $\Theta(x)=1$ for $x>0$ and zero otherwise.
This implies in particular
that the second and higher moments of $O_{\alpha\alpha}$ diverge.
We expect, from Eq.\,(\ref{eq:recursion}), below,  similar behaviour for $N>2$ and (provided $N>2$) for $O_{\alpha\beta}$
with $\alpha\not= \beta$.

In the remainder of this paper we sketch
our calculations and show how the results summarized
above can be generalized.
Calculations for the ensemble of Eq.\,(\ref{eq:ensemble})
can be done by extending the classical methods
of Dyson and Ginibre, while more general problems are most conveniently treated via ensemble-averaged resolvents, using the techniques of 
Refs.\,\cite{jan97,zee97,cha97}.

A direct computation of averages of $O_{\alpha\beta}$ involves
a $2N^2$-fold integration over the complex matrix
elements $J_{kl}$. The integral
is simplified considerably by changing variables
as described in \cite{meha}. 
Reducing $J$ by a unitary transformation, $U$, to upper triangular form,
so that $T\equiv U^{\dagger} J U$
has $T_{kl}=0$ for $k>l$,
we use as $N(N+1)$ coordinates, the real and imaginary parts of 
the non-zero elements of $T_{kl}$, 
and take the remaining coordinates from $U$ itself.
The required Jacobian is given by Mehta \cite{meha}.

In this basis, the diagonal elements of $T$ are the eigenvalues,  $T_{kk}=\lambda_k$. The first two pairs of eigenvectors are 
$|R_1\rangle = (1,0,\ldots,0)^\dagger$,
$\langle L_1| = (1,b_2,b_3,\ldots,b_N)$,
$|R_2\rangle = (-\bar{b}_2,1,0,\ldots,0)^\dagger$ and
$\langle L_2| = 
(0,1,d_3,\ldots,d_N)$, where the coefficients $b_l$ and $d_l$ are
determined by the recursion relations
\begin{equation}
\label{eq:recursion}
b_p = \frac{1}{\lambda_1-\lambda_p} \sum_{q=1}^{p-1} b_q T_{qp}\,,\hspace{0.4cm}
d_p = \frac{1}{\lambda_2-\lambda_p} \sum_{q=1}^{p-1} d_q T_{qp}\,,
\end{equation}
with $b_1 = 1$, $d_1 = 0$ and $d_2 = 1$. 
Correspondingly, the overlaps are 
\begin{eqnarray}
O_{11} &=&  \sum_{l=1}^N |b_l|^2\,,\\
O_{12} &=&  -\bar{b}_2 \sum_{l=1}^N b_l \bar{d}_l\,.
\end{eqnarray}
Performing the integrals over $U$ and $T_{kl}$ with $k<l$,
$O(z)$ and $O(z_1,z_2)$ can be expressed as averages with respect to
the joint probability density of the eigenvalues
 \begin{eqnarray}
 \label{eq:jpd}
  P(\lambda_1,..,\lambda_N) 
&\propto& \exp\big(\!-N\sum_{k=1}^N|\lambda_k|^2\big)\!\!\!\!
\prod_{1\leq i < j\leq N}\!\!\!\!\!|\lambda_i-\lambda_j|^2\,.
 \end{eqnarray}
Defining  $\langle \cdots\rangle_P$ as an average
with (\ref{eq:jpd}), we find
\begin{eqnarray}
\label{eq:P}
O(z_1) &=&\Big\langle \delta(z_1-\lambda_1)
\prod_{2\leq j\leq N}\Big(1+\frac{1}{N|\lambda_1-\lambda_j|^2}\Big)\Big\rangle_P
\end{eqnarray}
and
\begin{eqnarray}
\label{eq:Q}
\lefteqn{O(z_1,z_2) = -(N-1)
\Big\langle \delta(z_1-\lambda_1)\,\delta(z_2-\lambda_2)}\hspace{0.25cm}&&\\
&\times&
\frac{1}{N|\lambda_1-\lambda_2|^2}
\prod_{3\leq j \leq N} 
\Big(1+
\frac{1}{N(\lambda_1-\lambda_j)(\bar{\lambda}_2-\bar{\lambda}_j)}\Big)
\Big\rangle_P\,.
\nonumber
\end{eqnarray}
Performing the integrals over eigenvalues 
in Eqs.\,(\ref{eq:P}) and (\ref{eq:Q}), we obtain
explicit expressions in terms of $N\times N$ determinants.

For $z_1=0$, we are able to evaluate these determinants
in closed form by recursion, and to simplify the result further for $N\gg 1$.
For $z_1 \not= 0$ we are forced to take a less direct approach, 
which numerical tests show is a good approximation for finite $N$, and
which we can prove is exact in the limit $N \to \infty$.
We separate the contributions to each of the Eqs.\,(\ref{eq:P}) and (\ref{eq:Q}) into two factors: 
one from the $M$ eigenvalues closest to $z_1$, with $M\gg 1$, and another from the remaining eigenvalues. 
The first
can be evaluated using our result for $z_1=0$,
while the second can be calculated 
neglecting eigenvalue correlations, because its
fluctuations vanish as $M \to \infty$. Their combination is independent of $M$, for $M$ large, as it should be, 
and is as displayed in 
Eqs.\,(\ref{eq:largeNdiag}) -- (\ref{eq:np}).

An entirely different approach is necessary in order to treat other random matrix ensembles with ease, or to develop approximation schemes for spatially extended problems such as those of Refs.\,\cite{som88,haa92,hat96,cha97,nel98}. For these purposes we take as central objects the ensemble averages of products of the resolvents, $(z_1-J)^{-1}$ and $(\bar{z}_2-J^{\dagger})^{-1}$. As a demonstration, we examine a matrix ensemble with the probability distribution
\begin{eqnarray}
\label{eq:ensemble2}
\lefteqn{
P(J)\,dJ \propto}&&\hspace{2cm}\\
&&  \exp\Big(-\frac{N}{1-\tau^2}
\mbox{Tr}\left[J J^{\dagger}-\tau\,\mbox{Re}\,JJ\right]\Big)
\,\prod_{k,l=1}^N\!dJ_{kl}^\prime\,dJ_{kl}^{\prime\prime}\,,
\nonumber
\end{eqnarray}
with $\tau$ real and $-1\leq \tau \leq 1$. 
The non-zero cumulants are
$\langle J_{kl}\,\bar{J}_{kl}\rangle = 1/N$,
and $\langle J_{kl}\,\bar{J}_{lk}\rangle = \tau/N$.
This distribution, introduced in \cite{som88},
interpolates between the Gaussian 
unitary ensemble of Hermitian matrices, for $\tau=1$, Ginibre's ensemble 
(Eq.\,\ref{eq:ensemble}) for $\tau=0$, and complex antisymmetric 
matrices for $\tau=-1$.
In the limit $N\to \infty$, the eigenvalue density 
has the uniform value $d(z)=[\pi(1-\tau^2)]^{-1}$ within the ellipse defined by
$[\mbox{Re}\,z/(1-\tau)]^2+[\mbox{Im}\,z/(1+\tau)]^2 < 1$ 
and is zero elsewhere \cite{som88}.

We treat the ensemble (\ref{eq:ensemble2})
using the techniques described in \cite{jan97,zee97,cha97}.
These generate an expansion for $O(z_1,z_2)$ in powers of $(z_1-z_2)/N$,
and hence give $O(z_1,z_2)$ exactly in the limit $N\to \infty$, but supply information about $O(z)$ only indirectly, via the sum rule of Eq.\,\ref{eq:sumrule}. We start by
considering the $2N\times2N$ Hermitian
matrix
$\bbox{H} =\bbox{H}_0 + \bbox{H}_1$,
\begin{equation}
\label{eq:H}
\bbox{H}_0  = 
\left(
\begin{array}{cc}
\eta \,\,& \\  & -\eta
\end{array}
\right )\,,\hspace{0.4cm}
\bbox{H}_1  = 
\left(
\begin{array}{cc}
\,\,& z-J\\ \bar{z}-J^\dagger &
\end{array}
\right )
\end{equation}
with real $\eta > 0$, and its inverse 
\begin{equation}
\label{eq:greensfunction}
\bbox{G} = \bbox{H}^{-1}
=
\left(
\begin{array}{cc}
G_{11}\,\,& G_{12}\\G_{21} &G_{22}
\end{array}
\right ) \,.
\end{equation}
The resolvents are obtained taking $\eta \to 0$:
in this limit, $G_{21}=(z-J)^{-1}$ and $G_{12}=(\bar{z}-J^{\dagger})^{-1}$.
Expanding the Green function $\bbox{G}$ as a power series in
in $\bbox{H}_1$, its ensemble average can be written
\begin{equation}
\langle \bbox{G} \rangle = \bbox{G}_0
+ \bbox{G}_0 \bbox{\Sigma} \langle \bbox{G}\rangle
\end{equation}
where $\bbox{G}_0 = \bbox{H}_0^{-1}$ and
$\bbox{\Sigma}$ is a self-energy.  In the limit $N\to \infty$,
the self-consistent Born approximation for $\bbox{\Sigma}$
is exact \cite{jan97,zee97,cha97}, and the eigenvalue density can be obtained as
\begin{equation}
\label{eq:dos}
d(z) = \frac{1}{\pi} \frac{\partial}{\partial \bar{z}}
\lim_{\eta\rightarrow 0}
\Big\langle \frac{1}{N} \mbox{Tr}\, G_{21}(z)\Big\rangle\,.
\end{equation}
To study eigenvector correlations, it is necessary to calculate averages of products of $\bbox{G}$'s. In particular, the density
 \begin{equation}
 \label{eq:decomposition}
 D(z_1,z_2) = \delta(z_1-z_2) O(z_1) + O(z_1,z_2)
 \end{equation}
can be written as
\begin{eqnarray}
\label{eq:twopoint}
\lefteqn{
D(z_1,z_2) = \frac{1}{\pi^2}\frac{\partial}{\partial \bar{z}_1}
                            \frac{\partial}{\partial      {z}_2}}\hspace{1cm}&&\\
&\times&\lim_{\eta\rightarrow 0}
\Big\langle \frac{1}{N} \mbox{Tr}\; G_{21}(z_1)\,
G_{12}(z_2)\Big\rangle\,.
\nonumber
\end{eqnarray}
We therefore calculate $\bbox{R}(z_1,z_2)\equiv \langle\bbox{G}(z_1)\otimes\bar{\bbox{G}}(z_2)\rangle$, which obeys a Bethe-Salpeter equation \cite{jan97}: writing 
$\bbox{R}_0(z_1,z_2)
=\langle\bbox{G}_0(z_1)\rangle \otimes
 \langle\bar{\bbox{G}}_0(z_2)\rangle$,
\begin{equation}
\label{eq:bse}
\bbox{R} = \bbox{R}_0 + \bbox{R}_0 \bbox{\Gamma} \bbox{R}\,.
\end{equation}
In the limit $N\to \infty$ with $z_1 \not= z_2$, the vertex is simply $\bbox{\Gamma} = \mbox{diag}(1,\tau,\tau,1)$.
Solving Eq. (\ref{eq:bse}) for $\bbox{R}$, we obtain
\begin{eqnarray}
\label{eq:result2}
\lefteqn{D(z_1,z_2) = -\frac{1}{\pi^2|z_1-z_2|^4}}
\hspace{1cm}&&\\
&\times& \frac{(1-\tau^2)^2
-(1+\tau^2)z_1\bar{z}_2
+\tau(z_1^2 + \bar{z}_2^2)}{(1-\tau^2)}
\nonumber
\end{eqnarray}
for $z_1$ and $z_2$ within the support
of the density of states, and zero otherwise.
Since we have taken $z_1\not= z_2$, this is simply $O(z_1,z_2)$,
and for $\tau=0$, 
Eq. (\ref{eq:largeNoff}) is reproduced.
For $1-\tau \ll 1$, on the other hand,
$O(z_1,z_2)/[d(z_1)d(z_2)] \propto (1-\tau)$, so that
$O(z_1,z_2)/[d(z_1)d(z_2)]$  vanishes in the Hermitian
limit $\tau\rightarrow 1$, as expected. 

We thank B. Eckhardt and L. N. Trefethen for stimulating discussions.
This work was supported in part by EPSRC Grant GR/J78327,
and in part by SFB 393.

%
%

\end{multicols}

\begin{thebibliography}{10}
\bibitem{footnote}
Because of the cancelations implicit from Eq\,\ref{eq:sumrule}, Eq\,\ref{eq:ut2}
is most easily evaluated from the two-particle Green function, Eq\,33,
integrating $G_{21}(z_1)G_{12}(z_2)f_t(z_1)\bar{f}_t(z_2)$ around suitable contours
in the $z_1$ and $\bar{z}_2$ planes.
\end{thebibliography}
\end{document}